\documentclass[twocolumn,showpacs,preprintnumbers,amsmath,amssymb,superscriptaddress]{revtex4}

\usepackage{graphicx}
\usepackage{dcolumn}
\usepackage{bm}

\begin{document}

\preprint{APS/123-QED}

\title{Nuclear Magnetic Moment of the $^{57}$Cu Ground State}

\author{K. Minamisono}
\affiliation{National Superconducting Cyclotron Laboratory, Michigan State University, East Lansing, MI 48824, USA}

\author{P. F. Mantica}
\affiliation{National Superconducting Cyclotron Laboratory, Michigan State University, East Lansing, MI 48824, USA}
\affiliation{Department of Chemistry, Michigan State University, East Lansing, MI 48824, USA}

\author{T. J. Mertzimekis}
\affiliation{National Superconducting Cyclotron Laboratory, Michigan State University, East Lansing, MI 48824, USA}

\author{A. D. Davies}
\affiliation{National Superconducting Cyclotron Laboratory, Michigan State University, East Lansing, MI 48824, USA}
\affiliation{Department of Physics and Astronomy, Michigan State University, East Lansing, MI 48824, USA}

\author{M. Hass}
\affiliation{Department of Particle Physics, Weizmann Institute, 76100 Rehovot, Israel}

\author{J. Pereira}
\affiliation{National Superconducting Cyclotron Laboratory, Michigan State University, East Lansing, MI 48824, USA}

\author{J. S. Pinter}
\affiliation{National Superconducting Cyclotron Laboratory, Michigan State University, East Lansing, MI 48824, USA}
\affiliation{Department of Chemistry, Michigan State University, East Lansing, MI 48824, USA}

\author{W. F. Rogers}
\affiliation{Department of Physics, Westmont College, Santa Barbara, CA 93108, USA}

\author{J. B. Stoker}
\affiliation{National Superconducting Cyclotron Laboratory, Michigan State University, East Lansing, MI 48824, USA}
\affiliation{Department of Chemistry, Michigan State University, East Lansing, MI 48824, USA}

\author{B. E. Tomlin}
\affiliation{National Superconducting Cyclotron Laboratory, Michigan State University, East Lansing, MI 48824, USA}
\affiliation{Department of Chemistry, Michigan State University, East Lansing, MI 48824, USA}

\author{R. R. Weerasiri}
\affiliation{National Superconducting Cyclotron Laboratory, Michigan State University, East Lansing, MI 48824, USA}
\affiliation{Department of Chemistry, Michigan State University, East Lansing, MI 48824, USA}

\date{\today}

\begin{abstract}
The nuclear magnetic moment of the ground state of $^{57}$Cu($I^{\pi}$ = 3/2$^{-}$, $T_{\rm 1/2}$ = 196.3 ms) has been measured to be $|\mu(^{57}{\rm Cu})| = (2.00 \pm 0.05) \mu_{\rm N}$ using the $\beta$-NMR technique.  Together with the known magnetic moment of the mirror partner $^{57}$Ni, the spin expectation value was extracted as $\langle\Sigma\sigma_z\rangle$ = $-$0.78 $\pm$ 0.13.  This is the heaviest 
isospin $T$ = 1/2 mirror pair above the $^{40}$Ca region for which both ground state magnetic moments have been determined.  Discrepancy between present results and shell model calculations in full $fp$ shell giving $\mu(^{57}{\rm Cu}) \sim$ 2.4$\mu_{\rm N}$ and $\langle\Sigma\sigma_z\rangle \sim$ 0.5 implies significant shell breaking at $^{56}$Ni with the neutron number $N$ = 28.  
\end{abstract}

\maketitle

The magnetic moment is one of the important fundamental properties of the nucleus and provides key information to understand nuclear structure, especially ideas built on core polarization and the meson exchange effects in the shell model.  While the magnetic moments near the stability line in the nuclear chart are well known, we have only limited information on the magnetic moments far from the stability line.  Some of the challenges facing ground state nuclear moment measurements include the small spin polarization and low cross sections encountered in the production mechanism for nuclei far removed from the stability line.  It is important, however, to go far from the stability line in order to test predictive powers of theoretical models and to find possible variations of nuclear structure from the stability line.  

In this Letter, we report the first measurement of the magnetic moment of $^{57}$Cu.  Since $^{57}$Cu has one proton outside of the $^{56}$Ni core and the magnetic moment of $T$ = 1/2 mirror partner $^{57}$Ni is known, the current result completes the mirror pair and contributes important information for this shell region.  The difficulties in the small polarization and low cross section were overcome by using a proton pick up reaction of an intermediate-energy primary beam.  The small value for the $^{57}$Cu magnetic moment, discussed later in this paper, implies possible shell breaking at the neutron number $N$ = 28.

The magnetic moment can be expressed as the sum of isoscalar $\langle\Sigma\mu_0\rangle$ and isovector $\langle\Sigma\mu_z\rangle$ components as 
\begin{eqnarray}
\mu &=& \langle\Sigma \mu_0\rangle + \langle\Sigma \mu_z\rangle \\\nonumber
&=& \big\langle\Sigma \frac{l_z+(\mu_p+\mu_n)\sigma_z}{2}\big\rangle + \big\langle\Sigma \frac{\tau_z \{l_z+(\mu_p-\mu_n)\sigma_z\}}{2}\big\rangle,
\end{eqnarray}
where $l$ and $\sigma$ are the orbital and spin angular-momentum operators of the nucleon, respectively, $\tau$ is the isospin operator, $\mu_p$ and $\mu_n$ are the magnetic moments of free proton and neutron, respectively, and the sum is over all nucleons.  The isovector $\langle\Sigma\mu_z\rangle$ part is proportional to the isospin, $T_z$, and changes its sign for $T = \pm T_z$.  Provided that isospin is a good quantum number, the spin expectation value $\langle\Sigma\sigma_z\rangle$, which is a contribution from nucleon spins to the magnetic moment, can be extracted \cite{sugimoto} from the sum of mirror pair magnetic moments as
 \begin{equation}
 \langle\Sigma \sigma_z\rangle = \frac{\mu(+T_z)+\mu(-T_z)-I}{\mu_p+\mu_n-1/2},
 \end{equation}
where the total spin is $I = \langle\Sigma l_z\rangle + \langle\Sigma\sigma_z\rangle/2$.  In the $sd$ shell, all of the ground state magnetic moments of $T$ = 1/2 mirror nuclei have been measured and there remain several $T$ = 1 and 3/2 isospin multiplets to be measured.  There exists, however, virtually no information regarding similar $T$ = 1/2 pairs in the $fp$ shell.  A systematic trend of the spin expectation value depending on the shell structure has been observed \cite{sugimoto,hanna}, which gave firm understanding of the shell structure for $T$ = 1/2 mirror pairs.  
On the other hand, in the $fp$ shell only few mirror magnetic moments are known ($^{41}$Ca-$^{41}$Sc and $^{43}$Sc-$^{43}$Ti pairs \cite{raghavan}) and therefore it is essential to measure more magnetic moments in this region in order to explore the evolution of nuclear structure in the $fp$ shell and above.  Because $^{57}$Cu consists of the closed-shell $^{56}$Ni core plus one proton, $\mu(^{57}$Cu) is one of the most important among mirror nuclei in the $fp$ shell.

The long chain of  magnetic moments of odd-mass Cu isotopes has been measured over 2$p_{3/2}$, 1$f_{5/2}$ and 2$p_{1/2}$ neutron shells (from the mass $A$ = 59 through 69).  It was reported \cite{golovko} that a shell model calculation systematically overshoots the experimental data.  
The $\mu(^{59}$Cu) was measured \cite{golovko} to be even smaller than that of the shell model calculation and that of the prediction based on the systematics of local mirror pairs \cite{buck}.  
It is predicted that the $^{56}$Ni core is soft \cite{honma} and that any deviation from the shell model is a direct proof of a higher order configuration mixing and/or a shell breaking at $^{56}$Ni.  In this regard, the measurement of $\mu(^{57}$Cu) is essential because only one proton is outside the $^{56}$Ni closed core and the single particle contribution is expected to be strong.

The experiment was performed at the National Superconducting Cyclotron Laboratory (NSCL) at Michigan State University.  The $^{57}$Cu ions were produced from a primary beam of $^{58}$Ni accelerated up to 140 MeV/nucleon by the coupled cyclotrons and impinging on a 423 mg/cm$^2$ $^9$Be target.  A charge pick up reaction ($^{58}$Ni + p $\rightarrow$ $^{57}$Cu + 2n) was employed to produce $^{57}$Cu because a large nuclear polarization can be expected at the central momentum of the fragment \cite{groh}, where the production rate is higher, although the two-neutron evaporation may reduce the final nuclear polarization.  The primary beam was set at an angle of $+ 2^\circ$ relative to the normal beam axis at the production target in order to produce a polarized beam.  The $^{57}$Cu ions were separated from other reaction products by the A1900 fragment separator \cite{morrissey}.  The full angular acceptance ($\pm 2.5^\circ$) was used.  An achromatic wedge (300 mg/cm$^2$ Al) was placed at the second intermediate dispersive image for a separation of $^{57}$Cu based on relative energy loss in the wedge.  The central momentum of the $^{57}$Cu beam was chosen with a 1\% momentum acceptance to optimize the polarization.  The magnetic rigidity of the dipole magnets were $B\rho$ = 3.0372 and 2.6477 Tm for the first and second segments of the A1900, respectively.  A typical counting rate of $^{57}$Cu at the experimental port was about 350 particles/s/pnA and 5 pnA primary beam was available.  The major contaminations were $^{55}$Co and $^{56}$Ni.  However, because of their longer lifetimes and lower $\beta$-ray end-point energies, they did not impact the $\beta$-NMR measurement.  

The polarized $^{57}$Cu ions were delivered to the $\beta$-NMR apparatus \cite{mantica}, first passed in air and implanted into a single-crystal NaCl, which was 2 mm in thickness and tilted by 45$^{\circ}$ relative to the $\beta$-detector surface.  An external magnetic field of $H_0$ = 0.2 T was applied parallel to the direction of polarization to maintain the polarization in the crystal.  $^{57}$Cu decays to the levels of $^{57}$Ni emitting $\beta^+$ rays with a half life of 196.3 ms.  The branching ratio to the ground state ($I^{\pi}$ = 3/2$^-$) is 89.9\% and the maximum $\beta$-ray energy is 7.75 MeV with an asymmetry parameter of $A$ = +0.75 \cite{semon} and 8.6\% of $^{57}$Cu decays to the 1112.6 keV excited state ($I^{\pi}$ = 1/2$^-$).  The $\beta$ rays from stopped $^{57}$Cu were detected by a set of plastic-counter telescopes (double coincidence) placed 0$^\circ$(u) and 180$^\circ$(d) relative to the holding field direction.  Because of the asymmetric $\beta$-ray angular distribution from the polarized nuclei, $W(\theta) \sim 1+AP\cos{\theta}$, the counting rate between u and d counters is asymmetric depending on the polarization $P$, $A$ and the angle $\theta$ between the direction of $\beta$ ray and the polarization axis.  

\begin{figure*}
\includegraphics[width=6in,keepaspectratio,clip]{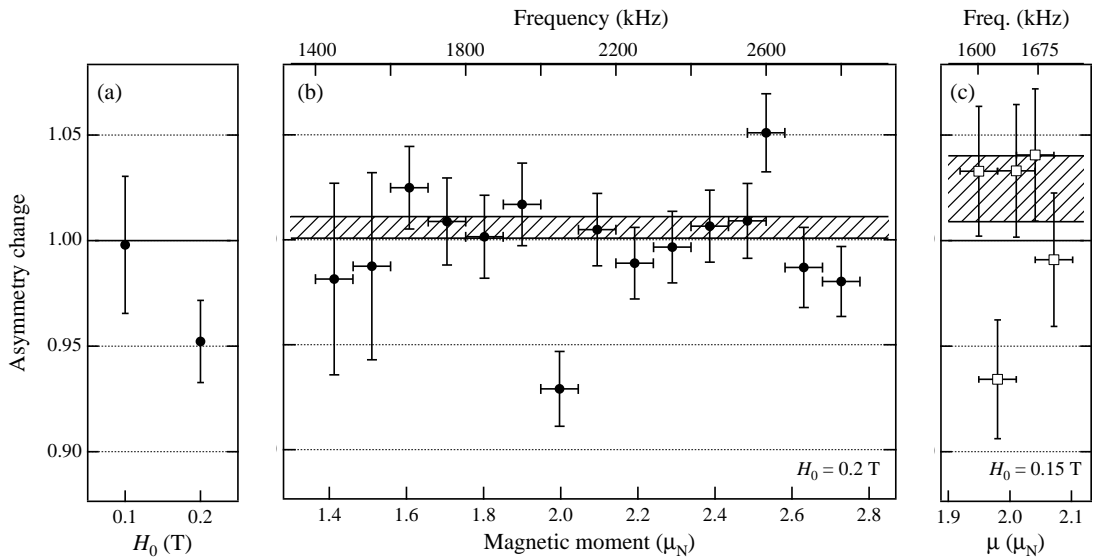}
\caption{\label{fig:NMRdata} Asymmetry change (1+2$AP$) of $^{57}$Cu in the NaCl as a function of $H_0$ is shown in (a).  In (b), the NMR data taken at 0.2 T with FM = $\pm$ 50 kHz is shown.  In (c), the data taken at 0.15 T with FM = $\pm$ 25 kHz is shown.  The hatched areas in (b) and (c) are base lines obtained from weighted sum of all the data except the resonance point.}
\end{figure*}

Prior to the NMR scan, an $H_0$ on and off technique \cite{anthony} was applied in order to deduce the polarization of $^{57}$Cu in NaCl.  The double ratio between up and down counters with (on) and without (off) $H_0$, $R = (N_u/N_d)_{\rm on}/(N_u/N_d)_{\rm off} \sim 1+2AP$, gives the asymmetry change, assuming that the polarization is not maintained in the NaCl without $H_0$.  However, an uncertainty in $R$ caused by the switching on and off of $H_0$ is introduced due to the effect of $H_0$ on the photomultipliers.  Therefore, the double ratio was normalized by the null polarization measurement, $R_{\theta = 2^{\circ}}/R_{\theta = 0^{\circ}}$, where the primary beam angle was set at  0$^\circ$.  Typically, $R_{\theta = 0^{\circ}}$ = 0.96(1) and 0.936(9) for $H_0$ = 0.1 and 0.2 T, respectively, were obtained.  As a result shown in Fig. \ref{fig:NMRdata} (a), an asymmetry change of about 0.95 was observed at $H_0$ = 0.2 T and no polarization at 0.1 T.  Considering the positive $A$ and assuming the near-side dominant reaction, the sign of the polarization is negative, where the direction defined by the momenta of the primary beam and the fragment, ${\bm k_i} \times {\bm k_f}$, is taken to be positive.

In the NMR measurement, a radio-frequency (rf) on and off technique \cite{mantica} with continuous $^{57}$Cu implantation into the NaCl was employed.  The frequency-modulated (FM) rf was applied to the $^{57}$Cu in NaCl every 60s through an rf coil, which made up an LCR circuit.  Typical FM, rf time to sweep the FM and the amplitude were $\pm$ 50 kHz, 20 ms and 0.3 mT, respectively.  The asymmetry change extracted from the ratio between rf off and on, $R = (N_u/N_d)_{\rm off}/(N_u/N_d)_{\rm on}$, was measured as a function of the applied frequency, $\nu$.  The $\mu$ is deduced from the relation $h\nu = \mu H_{\rm 0}/I$.   All the measurements were performed at room temperature and in the air.  


Several sets of NMR measurements were performed with different settings of $H_0$ and FM.  The NMR spectrum measured at $H_0$ = 0.2020 $\pm$ 0.0003 T with FM = $\pm$50 kHz is shown in Fig. \ref{fig:NMRdata} (b), where the asymmetry change is plotted as a function of the magnetic moment.  The applied frequency is also shown on the top axis.  A resonance was observed around $\mu$ = 2$\mu_{\rm N}$.  The amplitude and sign of the resonance are consistent with the result of the $H_0$ on/off measurement shown in Fig. \ref{fig:NMRdata} (a) within statistical error.  The hatched area is a base line obtained from weighted sum of all the data except the resonance point.  A positive peak around $\mu = 2.5\mu_{\rm N}$ was not considered a resonance because the sign of the asymmetry is different from the $H_0$ on/off.  Unfortunately the measurement at $H_0$ = 0.15 T with narrower FM =  $\pm$25 kHz was not conclusive as shown in Fig. \ref{fig:NMRdata} (c), which is due to the smaller polarization at 0.15 T expected from the result of the $H_0$ on/off shown in Fig. \ref{fig:NMRdata} (a) and the poorer statistics.  Thus only the results taken at $H_0$ = 0.2 T will be used in the analysis hereafter.  The observed asymmetry change at the resonance point was 0.929 $\pm$ 0.018, which is 3.9$\sigma$ number (99.99\% confidence level).  From the applied frequency of 2050 $\pm$ 50 kHz to the resonance point, the magnetic moment is obtained as 
\begin{equation}
|\mu(^{57}{\rm Cu})| = (2.00 \pm 0.05) \mu_{\rm N},
\end{equation}
where the error is from the FM only and the value is not corrected for the chemical shift, which is not known.  However, the chemical shifts of Cu in cuprous halides are known and they are about -1500 ppm \cite{lutz}, where a diamagnetic direction is taken to be positive ($\mu(^{57}$Cu) would shift by $-$0.003$\mu_{\rm N}$). 

\begin{figure}
\includegraphics[width=3.3in,clip,keepaspectratio]{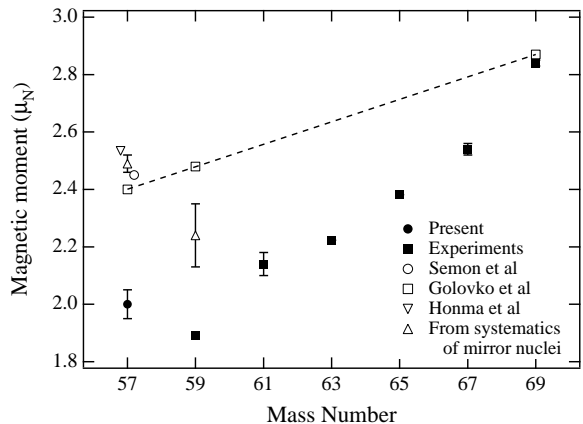}
\caption{\label{fig:cumoment} Magnetic moments of odd-mass Cu isotopes.  The solid circle is the present result and solid squares are the experimental data.  The dashed line and open squares are the theoretical calculation from \cite{golovko}, the open circle from \cite{semon}, and the open-inverse triangle from \cite{honma}.  The open triangles are predictions based on the systematics \cite{buck}.}
\end{figure}

The present result is shown in Fig. \ref{fig:cumoment} in solid circle together with the known magnetic moments of odd-mass Cu isotopes in solid squares in the same fashion as in Ref. \cite{golovko}.  The dotted line and open squares are the results from shell-model calculation in Ref. \cite{golovko}, where the $^{56}$Ni core is assumed to be principally a closed-shell nucleus and the authors include the effects of core polarization, meson exchange current, isobars, and relativistic corrections in perturbation theory.  For the residual interaction, the one-boson exchange potential with a short-range cut-off was used.  From the systematic deviation of $\mu$(Cu) from the calculation, a massive shell-breaking at $^{56}$Ni has been suggested \cite{golovko}.  The open circle is a theoretical calculation \cite{semon}, where 3p-2h excitation was allowed in the $fp$ shells and FPD6 effective interaction was used.  These calculations give values around $\mu(^{57}$Cu) = 2.45$\mu_{\rm N}$ and do not reproduce the present result.  In a calculation \cite{honma} using an effective interaction GXPF1 in the full $fp$ shell, 
the result is $\mu(^{57}$Cu) = +2.5345$\mu_{\rm N}$ with free nucleon $g$ factors, which is shown as open-inverse triangle in the Fig. \ref{fig:cumoment}.  If the effective $g$ factors are used, $g_s^{\rm eff}$ = 0.9$g_s^{\rm free}$, $g_l^{\rm eff}$ = 1.1 and $-$0.1 for protons and neutrons, respectively, which improves the agreement between experimental data of Ni and Zn magnetic moments in Ref. \cite{honma}, the result is $\mu(^{57}$Cu) = +2.4897$\mu_{\rm N}$.  Again the calculation gives a larger magnetic moment.  The open triangles in Fig. \ref{fig:cumoment} are predictions \cite{buck} based on systematics of known magnetic moment and of the $ft$-value of their mirror partners for the mass $A$ = 57 and 59 systems, respectively.  The experimental data are smaller but follow the trend of systematic values.  The experimental and theoretical magnetic moments of $^{57}$Cu and $^{57}$Ni are summarized in TABLE \ref{tab:table1}.  The agreement between experiment and theory for $^{57}$Cu is not good, whereas the agreement is good for the magnetic moment of mirror partner $^{57}$Ni.  The use of a smaller effective $g$ factor, for example $g_s^{\rm eff}$ = 0.75$g_s^{\rm free}$ used in the analysis of the $g$ factors of the first 2$^{+}$ excited state in Zn isotopes \cite{kenn}, gives closer value $\mu(^{57}$Cu) = +2.27$\mu_{\rm N}$ to the experiment, although the agreement for $\mu$($^{57}$Ni) becomes worse.  The considerably smaller $\mu(^{57}{\rm Cu})$ would imply significant shell breaking at $^{56}$Ni with $N$ = 28.

\begin{table}
\caption{\label{tab:table1} Magnetic moments of $^{57}$Cu-$^{57}$Ni and the spin expectation value of the mass $A$ = 57 system.}
\begin{ruledtabular}
\begin{tabular}{lccr}
 & $\mu$($^{57}$Cu)$\mu_{\rm N}$ & $\mu$($^{57}$Ni)$\mu_{\rm N}$ & $\langle\Sigma\sigma_z\rangle$ \\
\hline
exp\footnotemark[1] & 2.00 $\pm$ 0.05 & - & -0.78 $\pm$ 0.13\footnotemark[2] \\
exp\footnotemark[3] & - & -0.7975 $\pm$ 0.0014 & - \\
theor\footnotemark[4] & 2.48 & -0.71 & 0.71 \\
theor\footnotemark[5] & 2.40 & - & - \\
theor\footnotemark[6] & 2.49 & -0.79 & 0.51 \\
syst\footnotemark[7] & 2.49 $\pm$ 0.03 & - & 0.51 $\pm$ 0.08 \\
Schmidt & 3.79 & -1.91 & 1 \\
\end{tabular}
\end{ruledtabular}
\footnotetext[1]{Present work.}
\footnotetext[2]{From present work and Ref. \cite{ohtsubo}.}
\footnotetext[3]{From Ref. \cite{ohtsubo}.}
\footnotetext[4]{From Ref. \cite{semon}.}
\footnotetext[5]{From Ref. \cite{golovko}.}
\footnotetext[6]{From Ref. \cite{honma} with $g_s^{\rm eff}$ = 0.9$g_s^{\rm free}$, $g_l^{\rm eff}$ = 1.1 and -0.1 for proton and neutron, respectively.}
\footnotetext[7]{From Ref. \cite{buck} with $\mu$($^{57}$Ni) of Ref. \cite{ohtsubo}.}
\end{table}

Together with the known magnetic moment of the mirror partner $\mu(^{57}{\rm Ni}) = (-0.7975 \pm 0.0014)\mu_{\rm N}$ \cite{ohtsubo}, the spin expectation value is extracted as 
\begin{equation}
\langle\Sigma\sigma_z\rangle = -0.78 \pm 0.13,
\end{equation}
\begin{figure}
\includegraphics[width=3.2in,clip,keepaspectratio]{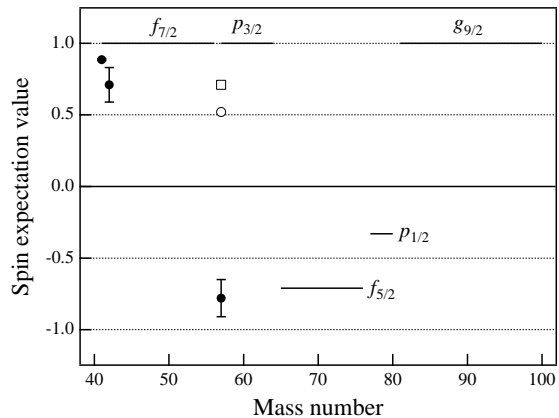}
\caption{\label{fig:spinexp} Spin expectation value of known $T$ = 1/2 pair in the $fpg$ shell.  The solid circles are the experimental data, open square is from \cite{semon} and the open circle from \cite{honma}.  The Schmidt values are indicated by the solid lines.}
\end{figure}
where the sign of $\mu(^{57}{\rm Cu})$ is assumed to be positive.  In TABLE \ref{tab:table1} the spin expectation values for the mass $A$ = 57 system are summarized and plotted in Fig. \ref{fig:spinexp}, where the experimental data are shown by the solid circles.  There are only three $T$ = 1/2 mirror pairs known in the $fp$ shell, namely $^{41}$Ca-$^{41}$Sc, $^{43}$Sc-$^{43}$Ti, and $^{57}$Ni-$^{57}$Cu pairs.  Theoretical calculations are also plotted for the $A$ = 57 from Ref. \cite{honma} and \cite{semon} in open circle and square, respectively.  The single-particle values (Schmidt values) are indicated by solid lines.  The smaller $\mu(^{57}$Cu) results in a large deviation and opposite sign in $\langle\Sigma\sigma_z\rangle$ from the shell model calculations.  Considering the systematic behavior of the $\langle\Sigma\sigma_z\rangle$ of $T$ = 1/2 nuclei in the $sd$ shell, where the values of $\langle\Sigma\sigma_z\rangle$ are restored close to the single-particle values around the shell closures, the present result indicates that $^{57}$Cu is at the middle of the shell evolution towards the shell closure and that there is a significant shell breaking at $^{56}$Ni with $N$ = 28.  Systematic studies of the spin expectation value towards the $N$ = 50 shell closure (1$g_{9/2}$) 
are of great interest.  

In summary, the magnetic moment of ground state $^{57}$Cu was measured for the first time by the $\beta$-NMR technique.  The $^{57}$Cu ions were produced from a 140 MeV/nucleon $^{58}$Ni beam at the NSCL/MSU impinging on a Be target and purified by the A1900 fragment separator.  A charge pick up reaction followed by two neutron evaporation was employed.  A large nuclear polarization, $P \sim -3\%$, at the central momentum of the fragment was obtained, where the $^{57}$Cu was implanted into a single-crystal NaCl.  In the NMR measurement, a resonance at an applied frequency of 2050 $\pm$ 50 kHz was observed, where the external magnetic field was 0.2020 $\pm$ 0.0003 T.  The obtained magnetic moment is $|\mu(^{57}{\rm Cu})|$ = 2.00 $\pm$ 0.05 $\mu_{\rm N}$, for which the chemical shift is not considered.  The shell model calculations available for $^{57}$Cu generally give $\mu(^{57}{\rm Cu})$ = 2.45 $\mu_{\rm N}$ and the agreement with the experiment is not good, whereas the experimental $\mu(^{57}{\rm Ni})$ is well reproduced.  The use of the effective $g^{\rm eff}$ factor makes the agreement better for $^{57}$Cu but not for  $^{57}$Cu and  $^{57}$Ni simultaneously.  The spin expectation value was extracted together with the known $\mu(^{57}{\rm Ni})$ as $\langle\Sigma \sigma_z\rangle$ = $-0.78 \pm 0.13$.  Again, the agreement with the theoretical value, $\langle\Sigma \sigma_z\rangle \sim$0.5, is not good.   The present results imply a significant shell breaking at $^{56}$Ni with $N$ = 28.

This work was supported in part by the National Science Foundation Grant PHY 01-10253 and PHY 99-83810.  The authors thank the NSCL operations staff for providing the beams for this experiment and thank B. A. Brown at the NSCL for valuable theoretical discussions.  One of the authors (MH) was supported in part by the Israel Science Foundation.

\end{document}